\documentstyle[preprint,aps]{revtex}
\begin{document}
\title{CONICAL SPACE-TIMES: A DISTRIBUTION THEORY APPROACH}
\author{F. Dahia \thanks{E-mail: fdahia@fisica.ufpb.br}  and C. Romero 
\thanks{E-mail: cromero@fisica.ufpb.br}\\Departamento de F\'{\i}sica\\
Universidade Federal da Para\'{\i}ba\\
C. Postal 5008 58059-970, J. Pessoa, Pb\\
Brazil}


\maketitle
\begin {abstract}

We consider the problem of calculating the Gaussian curvature of a 
conical 2-dimensional space by using concepts and techniques of 
distribution theory. We apply the results obtained to 
calculate the Riemannian curvature of the 4-dimensional conical space-time.
 We show that the method can be extended for calculating the curvature of
 a special class of more general space-times with conical singularity.

\end{abstract}

\newpage

\section{Introduction}
$         $ 

Although their great popularity in recent years may be mainly 
attributed to their close connection with cosmic strings \cite{Vilenkin}, 
conical space-times made their appearance in the physics literature 
by the end of the fifties through the work of Marder \cite{Marder}, whose 
interest was focused basically on topological aspects of locally 
isometric Riemannian spaces. A follow-up of Marder's seminal findings
 was to come some years later with an article by Sokolov and 
Starobinskii \cite{Sokolov} who, combining the celebrated Gauss-Bonet 
theorem of differential geometry and Einstein field equations, 
established a link between conical singularity and gravity. 
Just few years later conical space-times were rediscovered by 
Vilenkin \cite{Vilenkin} motivated by the investigation of gravitational 
effects of topological structures such as cosmic strings predicted 
by gauge theories \cite{Kibble}. Vilenkin's solution was found using 
the linear approximation of General Relativity. Then, Hiscock 
\cite{Hiscock} and Gott \cite{Gott}, followed later by Linet \cite{Linet}, 
worked out the exact solution by matching interior matter generated 
and exterior vacuum geometries, an approach which, in fact, had 
been developed with more generality by Israel in his attempts to 
characterize line sources in General Relativity \cite{Israel}.

In this paper we are concerned with the problem of how to 
evaluate the Riemann curvature tensor of a conical 4-dimensional space-time
 whose metric is known , thereby obtaining (via Einstein 
field equations) the energy-momentum tensor of the matter source. 
As in Sokolov's article, the specific form of the metric we consider
 reduces the problem to the calculation of the gaussian curvature or, 
equivalently, the curvature scalar of 2-dimensional 
spaces with conical singularity. However, rather 
than resorting to non-local concepts in order to circumvent 
singularities we make use of distribution theory to extend the 
concept of curvature. We show that at least for a number of cases 
this extension allows one to define curvature at points of the 
manifold where there is no tangent space. In the particular case of 
a conical space-time this approach reproduces in a very simple way 
the results obtained in the previous works mentioned already.

In the course of implementing the ideas above we became aware of other
 works which attack the problem of
 conical singularities \cite{Fursaev}. In particular, it should be mentioned 
the recent paper of Clark, Vickers and Wilson who apply Colombeau's generalized functions theory to calculate the distributional curvature of 
cosmic strings \cite{Clark}. On the other hand, topological defects and
 space-times with discontinuity in the derivatives of the metric tensor have 
been examined by Lichnerowicz, Israel, Taub and Letelier, among others 
\cite{Lich,Israel2,Taub,Letelier}. A quite general formulation of a 
mathematical framework 
to treat concentrated sources in General Relativity using distribution theory 
has been put forward by Geroch and Traschen \cite{Geroch}.
 However, these approaches differ from ours either from a  mathematical 
standpoint or in the degree of generality pursued by the authors.

\section{Preliminary concepts and definitions}
$         $

In this section we briefly review some elementary concepts of 
 distribution theory which will be used in extending the definition
 of curvature. For a clear and systematic treatment of distribution
 theory in Euclidian space the reader is referred to ref.\cite{Guelfand}.

To start with let us introduce some definitions. Consider a 2-dimensional
 manifold $S$ and a local coordinate system $(u,v)$. 
A $C^{\infty}$-scalar function $\varphi=\varphi(u,v)$ with
 compact support defined on $S$ is called a test-function. A continuous
 linear functional $F^{*}$, or  a {\it distribution}, is a rule which
associates a test-function $\varphi$ with a real number $(F^{*},\varphi)$
 such that the following conditions are satisfied:\\
$        $\\
i) $(F^{*},a_{1}\varphi_{1}+a_{2}\varphi_{2})=a_{1}(F^{*},\varphi_{1})
+a_{2}(F^{*},\varphi_{2})$, where $a_{1}$ and $a_{2}$ are real numbers 
(linearity condition);\\
$        $\\
ii) If the sequence of test-functions $\varphi_{1}, \varphi_{2}, ..., 
\varphi_{n}$ tends uniformly to zero, then the sequence of real numbers $(F^{*},
\varphi_{1}), (F^{*},\varphi_{2}), ..., (F^{*},\varphi_{n})$ approaches
 zero as well (continuity condition).\\
$   $

A scalar function $F=F(u,v)$ defined on $S$ is 
said to be locally integrable if 

\begin{eqnarray}
\int _{U}F(u,v)\varphi(u,v)\sqrt{g}dudv < \infty\nonumber
\end{eqnarray}
for any test-function $\varphi$ and an arbitrary 
compact domain $U \subset S$. Then, it is easy to see that any locally 
integrable function $F$ defines a distribution $F^{*}$ by the formula

\begin{equation}
\label{21}
(F^{*},\varphi)=\int_{U}F\varphi\sqrt{g}dudv,
\end{equation}
where $g$ denotes the determinant of the metric tensor $g_{ij}$ defined 
on $S$. At this point let us just note that due to $F$ and $\varphi$ 
being scalar 
functions the definition above is not coordinate-dependent. Any 
functional of the form (\ref{21}) is called a {\it regular distribution}. 
If a given distribution is not regular, i.e., if it cannot be put in the 
form (\ref{21}), it is called a {\it singular distribution}. The product of 
a given distribution $F^{*}$ by a scalar function $\alpha(x)\in C^{\infty}$
 is the functional $(\alpha F^{*})$ defined by 

\begin{equation}
\label{22}
((\alpha F^{*}),\varphi)=(F^{*},\alpha\varphi);
\end{equation}
whereas the derivative of $F^{*}$ with respect to the coordinate $u$ is 
the distribution $\left(\frac{\partial F}{\partial u}\right)^{*}$ given
 by the formula

\begin{equation}
\label{23}
\left(\left(\frac{\partial F}{\partial u}\right)^{*},\varphi\right)=-
\left(F^{*},\frac{1}{\sqrt{g}}\frac{\partial(\sqrt{g}\varphi)}
{\partial u}\right),
\end{equation}
where it must be assumed that the coordinate
 system is chosen such that $\sqrt{g}$ and $\frac{1}{\sqrt{g}}$ are also 
$C^{\infty}$-functions, a condition that in a sense may restrict the class of 2-dimensional 
manifolds upon which distributions are to be defined. (In fact, 
considerations concerning differentiability properties of $\sqrt{g}$ 
and $\frac{1}{\sqrt{g}}$ becomes relevants because we are 
broadening the ordinary 
definition of functional derivative in $\Re^{n}$ to include
 non-euclidian manifolds.)

Now let us consider geometry. One of the basic geometrical concepts when
 regarding 2-dimensional manifolds is the notion of Gaussian curvature 
$K$ \cite{Carmo}. It is a well known result that in two dimensions 
$K=\frac{R}{2}$ where $R$ is the curvature scalar. On the other 
hand if we are given the first quadratic form of a 2-dimensional 
manifold $S$

\begin{equation}
\label{24}
ds^{2}=Edu^{2}+Fdudv+Gdv^{2},
\end{equation}
where $E$, $F$ and $G$ are functions of the local coordinates $(u,v)$, 
then the curvature scalar $R$ can be calculated directly by the formula 
\cite{Carmo}

\begin{equation}
\label{25}
R=\frac{1}{\sqrt{g}}\left(\frac{\partial P}{\partial v}-\frac{\partial Q}
{\partial u}\right)
\end{equation}
where 

\begin{equation}
\label{26}
P\equiv\frac{1}{\sqrt{g}}\left(\frac{\partial F}{\partial u}-
\frac{\partial E}{\partial v}-\frac{1}{2}\frac{F}{E}\frac{\partial E}
{\partial u}\right),
\end{equation}

\begin{equation}
\label{27}
Q\equiv\frac{1}{\sqrt{g}}\left(\frac{\partial G}{\partial u}-\frac{1}
{2}\frac{F}{E}\frac{\partial E}{\partial v}\right),
\end{equation}
and $g=EG-\frac{1}{4}F^{2}$.

These geometrical definitions are all very well if the 2-dimensional manifold 
is what we call more properly a differentiable manifold. However, if there are
 points where the manifold is not smooth, and as a consequence no tangent 
space can be defined at these points, then the usual concept of 
curvature is meaningless. In such cases
 as, for example, the 2-dimensional cone, which is not a regular surface 
at the vertex,         
 equations like (\ref{25}) loses its applicability, and if we 
insist in defining curvature at points where the manifold is not regular
 we have to devise another definition outside  the scope of the usual 
differential geometry of surfaces. That is where distribution theory 
comes into play. Suppose that there exists a coordinate system in which 
$P$ and $Q$ are locally integrable functions.
 Then, we can define the curvature scalar functional by the following:

\begin{equation}
\label{28}
R^{*}=\frac{1}{\sqrt{g}}\left(\frac{\partial P^{*}}{\partial v}-
\frac{\partial Q^{*}}{\partial u}\right),
\end{equation}
where $P^{*}$ an $Q^{*}$ are the regular distributions constructed, 
respectively, from the functions $P$ and $Q$ according to the
 prescription (\ref{21}). At this point let us note that although the 
functions $P$ and $Q$ themselves are not scalars the combination in which 
they appear in equation (\ref{28}) behaves as a scalar. Therefore, applying 
$R^{*}$ to a test-function $\varphi$ we have

\begin{equation}
\label{29}
(R^{*},\varphi)=-\left(P^{*},\frac{1}{\sqrt{g}}
\frac{\partial\varphi}{\partial v}\right)+\left(Q^{*},\frac{1}{\sqrt{g}}
\frac{\partial\varphi}{\partial u}\right),
\end{equation}
whence

\begin{equation}
\label{210}
(R^{*},\varphi)=\int_{S}\left\{-P
\frac{\partial\varphi}{\partial v}+Q
\frac{\partial\varphi}{\partial u}\right\}dudv.
\end{equation}

Thus, given a 2-dimensional manifold with the metric tensor (\ref{24}) 
the equation (\ref{210}) written above may be considered as a definition 
of the curvature scalar regarded now as a functional or 
distribution. As we shall
 see in the next section, this extension of the concept of curvature will
 permit us to define and evaluate $R$ (or $K$) for a conical surface 
including its vertex.

\section{The curvature scalar of a conical surface}
$         $

In this section we apply the ideas developed previously to treat the 
problem of calculating the curvature scalar of the cone, the metric of 
which may be written in the form

\begin{equation}
\label{311}
ds^{2}=d\rho^{2}+\lambda^{2}\rho^{2}d\theta^{2},
\end{equation}
with $0\leq\rho <\infty$, $0\leq\rho<2\pi$ and $\lambda=$const$>0$. It is 
quite known that although (\ref{311}) leads to a vanishing curvature 
everywhere except for $\rho=0$, one cannot define a global coordinate 
system in which the metric tensor components are constants. The non-regular
 character of the conical space (\ref{311}) also manifests itself in that 
near the origin $g_{22}(\rho)=\lambda^{2}\rho^{2}$ does not fulfill the
 regularity conditions: $\sqrt{g_{22}}(\rho)\sim\rho$, $\frac{d\sqrt{g_{22}}
(\rho)}{d\rho}\sim1$ \cite{Linet,Eisenhart}. Also, regarding the cone as a surface 
embedded in $\Re^{3}$ a simple demostration that the cone is not a regular
 surface follows directly from the fact that it does not admit a
 differentiable parametrization in the neighborhood of the vertex \cite{Carmo}.  

Naturally, the conical space owes its name to the fact that its geometry 
may be identified with the geometry of a cone isometrically embedded in the
 3-dimensional euclidian space. The metric induced on the 
one-sheeted cone parametrized by the equation $z=a\rho$ may also be expressed, using 
cartesian coordinates, as 

\begin{equation}
\label{312}
ds^{2}=\left(1+\frac{a^{2}x^{2}}{x^{2}+y^{2}}\right)dx^{2}+
\left(1+\frac{a^{2}y^{2}}{x^{2}+y^{2}}\right)dy^{2}+
\frac{2a^{2}xy}{x^{2}+y^{2}}dxdy
\end{equation}
Before calculating the curvature scalar $R$ of the conical space as 
a functional we note that (\ref{311}) is not  
written in suitable coordinates as $\frac{1}{\sqrt{g}}$ is not $C^{\infty}$ 
everywhere. On the other hand, starting from (\ref{312}) one can 
check directly that $P$, $Q$, 
$\sqrt{g}$ and $\frac{1}{\sqrt{g}}$ all satisfy the conditions afore 
mentioned. Indeed, from (\ref{26}), (\ref{27}) and (\ref{312}) we have

\begin{equation}
\label{313}
P=\frac{2a^{2}}{\sqrt{1+a^{2}}}\left[\frac{y^{3}}{(x^{2}+y^{2})
(x^{2}+y^{2}+a^{2}x^{2})}\right],
\end{equation}

\begin{equation}
\label{314}
Q=-\frac{2a^{2}}{\sqrt{1+a^{2}}}\left[\frac{xy^{2}}{(x^{2}+y^{2})
(x^{2}+y^{2}+a^{2}x^{2})}\right],
\end{equation}
and $\sqrt{g}=\sqrt{1+a^{2}}$. At first sight, it might appear 
that the 
functions $P$ and $Q$ are not locally integrable as they are not 
bounded. 
That this is not so one can immediately see by going to the new
 coordinates defined by $x=r\cos\xi$, $y=r\sin\xi$ (in fact, 
the Jacobian of
 the transformation above regularize the singularity of $P$ 
and $Q$ at $r=0$, thereby leading to finite integrals).

Now, let us consider the integral (\ref{210}) which yields the curvature 
scalar as a distribution. We shall calculate this integral by first 
defining a small disc of radius $\epsilon$ with center at the vertex 
of the cone. We remove the disc from $S$ and call the remaining region
 $S_{\epsilon}$. Then, we have

\begin{equation}
\label{315}
(R^{*},\varphi)=\lim_{\epsilon\rightarrow 0}\int\int_{S_{\epsilon}}\left(-P
\frac{\partial\varphi}{\partial y}+Q
\frac{\partial\varphi}{\partial x}\right)dxdy.
\end{equation}
Clearly the legitimacy of the procedure above is 
guaranteed by the fact that the integrand is a locally integrable 
function.

Recalling that $\varphi$ has compact support and
 applying Green's theorem to the right-hand side of (\ref{315}) we obtain

\begin{equation}
\label{316}
(R^{*},\varphi)=\lim_{\epsilon\rightarrow 0}\left[\int_{S_{\epsilon}}
\left(
\frac{\partial P}{\partial y}-\frac{\partial Q}{\partial x}\right)
\varphi dxdy-
\int_{\partial S_{\epsilon}}(Pdx + Qdy)\varphi\right], 
\end{equation}
where $\partial S_{\epsilon}$ denotes the boundary of $S_{\epsilon}$ and the integration is performed in the anticlockwise sense.

A quick look at eq. (\ref{316}) will reveal us the presence of a term 
proportional to the curvature scalar in the integrand of the surface 
integral (see eq. (\ref{25})). As for the line integral let us find out 
its geometrical meaning. With this purpose let us consider the covariant 
derivative of the vector $\stackrel{\wedge}{e}_{u}=\frac{1}{\sqrt{E}}
\partial_{u}$ along the curve $\gamma$ defined parametrically by 
$\gamma(s)=(u(s),v(s))$. We have

\begin{equation}
\label{317}
\frac{D\stackrel{\wedge}{e}_{u}}{Ds}=\frac{d}{ds}\left(\frac{1}{\sqrt{E}}
\right)\partial_{u}+\frac{1}{\sqrt{E}}\Gamma_{\nu u}^{\mu}
\left(\frac{d\gamma}{ds}\right)^{\nu}\partial_{\mu},
\end{equation}
the indices $(\mu,\nu)$ running through $u$ and $v$. Let us 
define the vector

\begin{equation}
\label{318}
\stackrel{\wedge}{e}_{u}^{\perp}=\frac{\stackrel{\wedge}{e}_{v}-
<\stackrel{\wedge}{e}_{u},\stackrel{\wedge}{e}_{v}>\stackrel{\wedge}{e}_{u}}
{\parallel\stackrel{\wedge}{e}_{v}-<\stackrel{\wedge}{e}_{u},\stackrel{\wedge}{e}_{v}>\stackrel{\wedge}{e}_{u}\parallel}=
\frac{\sqrt{E}}{\sqrt{g}}\left(\partial_{v}-\frac{1}{2}\frac{F}{E}\partial_{u}\right),
\end{equation}
where $\stackrel{\wedge}{e}_{v}=\frac{1}{\sqrt{G}}\partial_{v}$ and 
the symbols $\parallel$  $\parallel$, $<$, $>$ denote norm and inner product, 
respectively. It is clear that the pair $\{\stackrel{\wedge}{e}_{u},\stackrel
{\wedge}{e}_{u}^{\perp}\}$ constitute a positive vector basis provided that 
$\{\stackrel{\wedge}{e}_{u},\stackrel{\wedge}{e}_{v}\}$ is positive. 
The projection 
of the covariant derivative $\frac{D\stackrel{\wedge}{e}_{u}}{Ds}$ 
onto the orthogonal vector $\stackrel{\wedge}{e}_{u}^{\perp}$ is called the 
algebraic value \cite{Carmo} of the derivative and is denoted by 
$\left[\frac{D\stackrel{\wedge}{e}_{u}}{Ds}\right]$, i.e., 

\begin{eqnarray}
\left[\frac{D\stackrel{\wedge}{e}_{u}}{Ds}\right]\equiv
\left<\frac{D\stackrel{\wedge}{e}_{u}}{Ds},\stackrel{\wedge}{e}_{u}^{\perp}
\right>.\nonumber
\end{eqnarray}

From (\ref{317}) and (\ref{318}) it follows that

\begin{equation}
\label{319}
\left[\frac{D\stackrel{\wedge}{e}_{u}}{Ds}\right]=
\frac{\sqrt{g}}{E}\Gamma_{u\lambda}^{v}\left(\frac{d\gamma}{ds}\right)
^{\lambda}.
\end{equation}

Since we have

\begin{equation}
\label{320}
\Gamma_{uu}^{v}=\frac{1}{g}
\left(-\frac{1}{4}F\frac{\partial E}{\partial u}+
\frac{1}{2}E\frac{\partial F}{\partial u}-
\frac{1}{2}E\frac{\partial E}{\partial v}\right),
\end{equation}
and

\begin{equation}
\label{321}
\Gamma_{uv}^{v}=\frac{1}{g}
\left(-\frac{1}{4}F\frac{\partial E}{\partial v}+
\frac{1}{2}E\frac{\partial G}{\partial u}\right),
\end{equation}
we are led to the equation

\begin{equation}
\label{322}
\left[\frac{D\stackrel{\wedge}{e}_{u}}{Ds}\right]=
\frac{1}{2}
\left(P\frac{du}{ds}+Q
\frac{dv}{ds}\right).
\end{equation}
Therefore, (\ref{315}) has the form

\begin{equation}
\label{323}
(R^{*},\varphi)=\lim_{\epsilon\rightarrow 0}\left[\int_{S_{\epsilon}}
\left(
\frac{\partial P}{\partial y}-\frac{\partial Q}{\partial x}\right)
\varphi dxdy-
2\int_{\partial S_{\epsilon}}\left[\frac{D\stackrel{\wedge}{e}_{x}}
{Ds}\right]\varphi ds\right], 
\end{equation}
where the geometrical meaning of the integrand of the line integral is 
expli-citly displayed. To get further insight into the concept of the algebraic
 value of the covariant derivative of a given unit vector $\stackrel
{\wedge}{\omega}$, let us make use of the following result \cite{Carmo}: 
if $\chi$ is the angle between any two unit vectors $\stackrel{\wedge}
{\omega}$ and $\stackrel{\wedge}{z}$, both defined along a certain 
curve $\lambda(s)$ and $\chi$ is taken from $\stackrel{\wedge}{z}$ to 
$\stackrel{\wedge}{\omega}$, then we have 

\begin{equation}
\label{324}
\left[\frac{D\stackrel{\wedge}{\omega}}{Ds}\right]=
\left[\frac{D\stackrel{\wedge}{z}}{Ds}\right]+\frac{d\chi}{ds}
\end{equation}
Let us suppose that the vector field $\stackrel{\wedge}{z}$ is 
constructed by parallel-transporting it along the curve $\gamma$. In 
this case $\frac{D\stackrel{\wedge}{z}}{Ds}=0$, hence $\left[\frac{D\stackrel
{\wedge}{\omega}}{Ds}\right]=\frac{d\chi}{ds}$. Thus, the algebraic value of 
the covariant derivative of a unit vector $\stackrel{\wedge}{\omega}$ 
may be regarded as a measure of the variation of the angle between 
$\stackrel{\wedge}{\omega}$ and a parallel transported vector 
$\stackrel{\wedge}{z}$. 

After all these considerations (\ref{323}) takes the form

\begin{equation}
\label{326}
(R^{*},\varphi)=\lim_{\epsilon\rightarrow 0}\left[\int_{S_{\epsilon}}
R_{reg}\varphi dxdy+
2\int_{\partial S_{\epsilon}}\frac{d\chi}{ds}\varphi ds\right], 
\end{equation}
where we have written $R_{reg}$ to highlight the fact that $R$ is 
calculated in a region where the conical surface is regular, and 
$\chi$ denotes the angle from the vector $\stackrel{\wedge}{e_{x}}$ to  a unit vector $\stackrel{\wedge} 
{z}$ parallel-transported along $\partial S_{\epsilon}$.

It is apparent that the first term of the right-hand side of the 
equation above vanishes since $R=0$ everywhere except at the origin. Thus, 
we have

\begin{eqnarray}
\label{327}
(R^{*},\varphi)&=&\lim_{\epsilon\rightarrow 0}\hbox{  }
2\int_{\partial S_{\epsilon}}\frac{d\chi}{ds}\varphi ds
\nonumber \\
&=& 2\varphi(0)\lim_{\epsilon\rightarrow 0}\int_{\partial S_{\epsilon}}
\frac{d\chi}{ds} ds,
\end{eqnarray}
the last step being justified by a known theorem concerning real continuous 
functions (see appendix). 

The limit in $(26)$ can be  calculated immediately if we note that

\begin{eqnarray}
\label{328}
\int_{\partial S_{\epsilon}}
\frac{d\chi}{ds} ds=\chi(s_{f})-\chi(s_{i}),
\end{eqnarray}
where $s_{f}$ and $s_{i}$ are the final and  initial values of the parameter 
$s$ on $\partial S_{\epsilon}$. Evidently, $s_{f}$ and $s_{i}$ represent 
the same point on $S_{\epsilon}$ since $\partial S_{\epsilon}$ is a 
circumference,  so the final and initial points coincide.

Further, by the  definition of the angle $\chi$, we know that 
$\chi(s_{f})$ is the angle between $\stackrel{\wedge}{z}(s_{f})$ 
and $\stackrel{\wedge}
{e_{x}}(s_{f})$, i.e., between the vector parallel-transported and the vector 
$\stackrel{\wedge}{e_{x}}$, both taken at $s_{f}$ on $\partial S_{\epsilon}$. 
Analogously, $\chi(s_{i})$ measures the angle between $\stackrel{\wedge}
{z}(s_{i})$ and $\stackrel{\wedge}{e_{x}}(s_{i})$. Since the endpoints 
of $\partial S_{\epsilon}$ coincide and $\stackrel{\wedge}{e_{x}}$ is a 
vector field, we have $\stackrel{\wedge}{e_{x}}(s_{f})= \stackrel{\wedge}
{e_{x}}(s_{i})$. Thus $\triangle\chi\equiv\chi(s_{f})-\chi(s_{i})$ is indeed 
the angle between $\stackrel{\wedge}{z}(s_{f})$ and $\stackrel{\wedge}
{z}(s_{i})$, i.e., the angular deviation of the vector that has been 
parallel-transported along $\partial S_{\epsilon}$. Clearly, $\Delta\chi$ is 
a quantity which depends on the global properties of the manifold 
$S$. In this way we see that the singular term of $R^{*}$ is related to 
the topology of the manifold, not only to its geometry.

In the cone case, this term $\Delta\chi$ is equal to the angular deficit 
  $\Delta\chi=2\pi(1-\lambda)$. Then, from $(26)$, 
we are led to the result:

\begin{equation}
\label{330}
(R^{*},\varphi)=2\varphi(0)[2\pi(1-\lambda)]=4\pi(1-\lambda)\varphi(0),
\end{equation}
which may be expressed in  more familiar form in terms of Dirac delta function as 

\begin{equation}
\label{331}
R^{*}=4\pi(1-\lambda)\delta^{(2)}(\rho),
\end{equation}
where, by definition, 

\begin{eqnarray}
\int{\delta^{(2)}(\rho)\sqrt{g}d\rho d\theta}=1.\nonumber
\end{eqnarray}

\section{Application to the generalized cone}
$         $

We can easily generalize the results obtained in the previous section by 
considering a surface embedded in the 3-dimensional Euclidean surface 
$\Sigma$ defined by the equation $z=\alpha(\rho)$, where $\alpha$ is an 
arbitrary $C^{\infty}-$function. The induced metric in $\Sigma$ is given 
by the line element

\begin{equation}
\label{4032}
ds^{2}=[1+\alpha^{'2}]d\rho^{2}+\rho^{2}d\theta^{2},
\end{equation}
where prime denotes derivative with respect to $\rho$. In cartesian 
coordinates, the equation above may be written as

\begin{equation}
\label{4033}
ds^{2}=\stackrel{\sim}{E}dx^{2}+\stackrel{\sim}{F}dxdy\stackrel{\sim}{G}
dy^{2},                         
\end{equation}
with $\stackrel{\sim}{E}=1+\frac{\alpha^{'2}x^{2}}{x^{2}+y^{2}},
\stackrel{\sim}{F}=2\frac{\alpha^{'2}xy}{x^{2}+y^{2}}$ and 
$\stackrel{\sim}{G}=1+\frac{\alpha^{'2}y^{2}}{x^{2}+y^{2}}$.
As we already know, the curvature scalar corresponding to (\ref{4033}) is 
obtained from (\ref{25}) with $\sqrt{g}=\sqrt{1+\alpha^{'2}}$ and the 
functions $\stackrel{\sim}{P}$ and $\stackrel{\sim}{Q}$ given as below:

\begin{equation}
\label{4034}
\stackrel{\sim}{P}(x,y)=P_{[\alpha^{'}]}+\bar{P},
\end{equation}

\begin{equation}
\label{4035}
\stackrel{\sim}{Q}(x,y)=Q_{[\alpha^{'}]}+\bar{Q},
\end{equation}
where

\begin{equation}
\label{4036}
P_{[\alpha^{'}]}=\frac{2\alpha ^{'2}}{\sqrt{1+\alpha ^{'2}}}
\frac{y^{3}}{[(1+\alpha ^{'2})x^{2}+y^{2}](x^{2}+y^{2})},
\end{equation}

\begin{equation}
\label{4037}
\bar{P}=\frac{2\alpha^{'}\alpha^{''}}{\sqrt{1+\alpha^{'2}}}
\frac{x^{2}y}{\sqrt{x^{2}+y^{2}}
[(1+\alpha ^{'2})x^{2}+y^{2}]},
\end{equation}

\begin{equation}
\label{4038}
Q_{[\alpha^{'}]}=-\frac{2\alpha ^{'2}}{\sqrt{1+\alpha ^{'2}}}
\frac{y^{3}}{[(1+\alpha ^{'2})x^{2}+y^{2}](x^{2}+y^{2})},
\end{equation}
and

\begin{equation}
\label{4039}
\bar{Q}=\frac{2\alpha^{'}\alpha ^{''}}{\sqrt{1+\alpha ^{'2}}}
\frac{xy^{2}}{\sqrt{x^{2}+y^{2}}
[(1+\alpha^{'2})x^{2}+y^{2}]}.
\end{equation}

At this point two comments are in order. The first is that, as we shall 
see later, the functions $P_{[\alpha ']}$ and $Q_{[\alpha ']}$ 
are recognized as the part of the curvature which accounts for the conical 
singularity. (Note that $P_{[\alpha ']}$ and $Q_{[\alpha ']}$ reduce 
to (\ref{313}) and ({\ref{314}), respectively, in the particular case 
$\alpha(\rho)=a\rho$.) The second comment concerns the requirements which must 
be fulfilled by the function $\alpha(\rho)$: for the curvature scalar to be 
well defined  as a functional we must assure ourselves that i)$\sqrt{g}$ is 
$C^{\infty}$ and ii) $\stackrel{\sim}{P}$ and $\stackrel{\sim}{Q}$ are locally 
integrable. Then, we define the curvature scalar functional by

\begin{equation}
\label{4040}
(R^{*},\varphi)=\lim_{\epsilon\rightarrow 0}
\int_{\partial S_{\epsilon}}\left(\frac{\partial{\stackrel{\sim}{P}}}
{\partial y}-
\frac{\partial{\stackrel{\sim}{Q}}}{\partial x}\right)\varphi dxdy-
\int_{\partial S_{\epsilon}}(Pdx+Qdy)\varphi,
\end{equation}
where, the region $S_{\epsilon}$ and its boundary $\partial S_{\epsilon}$ are
defined as before.

Considering the equation above it is readily seen that $\bar{P}$ and $\bar{Q}$ 
do not contribute to the line integral. Indeed, parametrizing 
$\partial S_{\epsilon}$ by ($x=\varepsilon\cos\theta$, $y=\epsilon\sin\theta$) it 
is straightforword to see that on the boundary $\partial S_{\epsilon}$

\begin{equation}
\label{4041}
\bar{P}dx+\bar{Q}dy=0
\end{equation}
Thus, the line integral which appears in (\ref{4040}) reduces to 

\begin{equation}
\label{4042}
\int_{\partial S_{\epsilon}}(P_{[\alpha^{'}]}dx+Q_{[\alpha^{'}]}dy)\varphi,
\end{equation}

To compute this term we proceed exactly as we did for the cone case. In this way, one can simply show that,

\begin{equation}
\label{4043}
\int_{\partial S_{\epsilon}}(P_{[\alpha^{'}]}dx+Q_{[\alpha^{'}]}dy)=
4\pi(\lambda(\epsilon)-1),
\end{equation}
where $\lambda(\epsilon)=\frac{1}{\sqrt{1+\alpha^{'2}(\epsilon)}}$; whence it follows that

\begin{equation}
\label{4044}
\lim_{\epsilon\rightarrow 0}
\int_{\partial S_{\epsilon}}(P(\alpha^{'})dx+Q(\alpha^{'})dy)\varphi=
4\pi(\lambda(0)-1)\varphi(0).
\end{equation}

On the other hand, similarly to the cone case, if we assume that $\Sigma$ is regular at $\rho\neq 0$, the surface integral term of the equation (\ref{4040}) 
may be put in the form

\[\int_{S_{\epsilon}}R_{reg}\varphi dxdy,\] where
\[R_{reg}=\frac{2\alpha^{'}\alpha^{''}}{(1+\alpha^{'2})^{2}\rho}.\]

Therefore, we finally conclude that the curvature scalar defined as a distribution will be given by

\begin{equation}
\label{4045}
R^{*}=\frac{2\alpha^{'}\alpha^{''}}{(1+\alpha^{'2})^{2}\rho}
+4\pi(1-\lambda)\delta^{(2)}(\rho).
\end{equation} 

According to (\ref{4032}) if we have a conical singularity at $\rho =0$, then by definition, $\alpha^{'}(0)\neq 0$. In this case, we see that the distribution $R^{*}$ contains a singular part which appears as a delta function.

\section{Extension to 4-dimensional space-time}
$         $

In the previous section we have shown by using techniques imported 
from distribution 
theory how to define the curvature scalar of a 2-dimensional
 ma-nifold at points where conical singularities exist. In particular, we 
have worked out the case of the 2-dimensional cone by
 calculating explicitly its curvature 
scalar as a distribution. It turns out, as we shall see, that the same
 mathematical treatment may be applied to calculate the Riemannian 
curvature of a 4-dimensional manifold $M$ which also has conical 
singularities provided that $M$ near the singularity admits a 
suitable coordinate system in which its line element has the special form

\begin{equation}
\label{432}
ds^{2}=(Adt^{2}+Bdtdz+Cdz^{2})+(Edu^{2}+Fdudv+Gdv^{2}),
\end{equation}
where the metric components $(A,B,C)$ depend solely on $(t,z)$ whereas 
$(E,F,G)$ are functions of $(u,v)$ only.

In fact, this was the case considered by Sokolov \cite{Sokolov} in which the 
conical 4-dimensional space-time $M$ may be regarded as the direct 
product $\Re^{2}\bigotimes Q^{2}$, $Q^{2}$ standing for the conical surface. 
Truly, the possibility of decomposing a 4-dimensional space-time $M$ 
into 2-dimensional submanifolds $M_{1}$ and $M_{2}$ greatly simplifies
 the calculation of the Riemannian curvature of $M$ allowing, as a 
consequence, the use of distribution theory to treat the Einstein tensor
 as a functional. To see this, one can easily verify that the Christoffell
 symbols $\Gamma_{\mu\nu}^{\lambda}$ of $M$ calculated directly from 
(\ref{432}) have the peculiar form 

\begin{eqnarray}
\Gamma_{\mu\nu}^{\lambda}=\left\{ \begin{array}{ll}
^{(1)}\Gamma_{\mu\nu}^{\lambda}, \mbox{ for } \lambda, \mu, \nu= t \mbox{ or } z;\\
^{(2)}\Gamma_{\mu\nu}^{\lambda}, \mbox{ for } \lambda, \mu, \nu= u \mbox{ or } v;\\
0, \mbox{ otherwise};\end{array}\right.\nonumber
\end{eqnarray}
where $^{(1)}\Gamma_{\mu\nu}^{\lambda}(t,z)$ and $^{(2)}\Gamma_{\mu\nu}
^{\lambda}(u,v)$ are the Christoffell symbols of the submanifolds $M_{1}$ 
and $M_{2}$, calculated from the metrics $^{(1)}ds^{2}=Adt^{2}+Bdtdz
+Cdz^{2}$ and $^{(2)}ds^{2}=Edu^{2}+Fdudv+Gdv^{2}$, respectively. 
Analogously, the components of the Riemann tensor are also decomposed
 into separate parts, as below:

\begin{eqnarray}
R_{\mu\nu\kappa}^{\lambda}=\left\{ \begin{array}{ll}
^{(1)}R_{\mu\nu\kappa}^{\lambda}, \mbox{ for } \lambda, \mu, \nu, \kappa=
 t \mbox{ or } z;\\
^{(2)}R_{\mu\nu\kappa}^{\lambda}, \mbox{ for } \lambda, \mu, \nu, \kappa=
 u \mbox{ or } v;\\
0, \mbox{ otherwise};\end{array}\right.\nonumber
\end{eqnarray}
where, as before, $^{(1)}R_{\mu\nu\kappa}^{\lambda}(t,z)$ and $^{(2)}
R_{\mu\nu\kappa}^{\lambda}(u,v)$ refer to the Riemannian curvature 
tensors of $M_{1}$ and $M_{2}$, respectively.

Further simplification is achieved by noting that due to a distinctive 
property of two dimensions, the Riemann tensors of the submanifolds 
may be written as \cite{Weinberg}

\begin{equation}
\label{433}
^{(i)}R_{\mu\nu\lambda\kappa}=\frac{1}{2}(g_{\mu\lambda}^{(i)}
g_{\nu\kappa}^{(i)}-g_{\mu\kappa}^{(i)}g_{\nu\lambda}^{(i)})^{(i)}R,
\end{equation}
where the index $i=1, 2$ refers, evidently, to geometric quantities 
defined on $M_{1}$ and $M_{2}$. From (\ref{433}) we calculate the Ricci 
tensor of $M$, which takes the form

\begin{eqnarray}
R^{\lambda}_{\mu}=\left\{ \begin{array}{ll}
\frac{1}{2}^{(1)}R\delta_{\mu}^{\lambda}, \mbox{ for } \lambda, 
\mu= t \mbox{ or } z;\\
\frac{1}{2}^{(2)}R\delta_{\mu}^{\lambda}, \mbox{ for } \lambda, 
\mu= u \mbox{ or } v;\\
0, \mbox{ otherwise};\end{array}\right.\nonumber
\end{eqnarray}
Contracting the indices $\lambda$ and $\mu$ we have

\begin{eqnarray}
 R=^{(1)}R+^{(2)}R,\nonumber
\end{eqnarray}
which leads to the following expression for the Einstein tensor:

\begin{eqnarray}
G^{\lambda}_{\mu}=\left\{ \begin{array}{ll}
-\frac{1}{2}^{(2)}R\delta_{\mu}^{\lambda}, \mbox{ for } \lambda, \mu= t \mbox{ or } z;\\
-\frac{1}{2}^{(1)}R\delta_{\mu}^{\lambda}, \mbox{ for } \lambda, \mu= u \mbox{ or } v;\\
0, \mbox{ otherwise};\end{array}\right.\nonumber
\end{eqnarray}
(Just note that in this case an inversion in the position of the indices
 has occurred: the components corresponding to one set of coordinates 
now are functions of the other set.)

Accordingly, it is natural to define the functional $(G^{\lambda}_{\mu})
^{*}$ corresponding to the mixed components of the Einstein tensor by

\begin{eqnarray}
{G^{\lambda}_{\mu}}^{*}=\left\{ \begin{array}{ll}
-\frac{1}{2}^{(2)}R^{*}\delta_{\mu}^{\lambda}, \mbox{ for } \lambda, \mu= t \mbox{ or } z;\\
-\frac{1}{2}^{(1)}R^{*}\delta_{\mu}^{\lambda}, \mbox{ for } \lambda, \mu= u \mbox{ or } v;\\
0, \mbox{ otherwise};\end{array}\right.\nonumber
\end{eqnarray}
where the functionals $^{(1)}R^{*}$ and $^{(2)}R^{*}$ are defined as
 in equation (\ref{210}).

In this way we have found that if the metric of a 4-dimensional manifold
 $M=M_{1}\bigotimes M_{2}$ can be written in the special form (\ref{432}),
 then the Einstein tensor of $M$ is directly obtainable from the curvature
 scalars of the submanifolds $M_{1}$ and $M_{2}$. If the manifold $M$ is 
not regular everywhere, then such non-regularity may manifest itself as
 a non-regularity of one of (or both) its 2-dimensional 
submanifolds. In this situation, as we have shown earlier, the 
problem of calculating the Riemannian curvature of $M$ is amenable to 
a distribution theory approach.

To conclude the section it is worth noting that if we make a coordinate 
transformation of the type $t^{'}=t^{'}(t,z)$, $z^{'}=z^{'}(t,z)$, $u^{'}=
u^{'}(u,v)$ and $v^{'}=v^{'}(u,v)$, then the separable form of (\ref{432}) 
is preserved. In this case, the components $G^{\lambda}_{\mu}$ of the 
Einstein tensor do not change, i.e., they are invariant. Naturally, in 
this new coordinates $G_{\mu}^{\lambda}$ may still be defined as a 
functional provided the following is also  preserved by the coordinates 
transformation: i) the new functions $P_{i}$ and $Q_{i}$, as defined in 
 (\ref{26}) and (\ref{27}), are locally 
integrable in each submanifold $M_{i} (i=1,2)$; ii) 
$\sqrt{\stackrel{(2)}{ }g_{i}}$ and $\frac{1}
{\sqrt{\stackrel{(2)}{ }g_{i}}}$ are $C^{\infty}$ functions, where by 
$\stackrel{(2)}{ }g_{i}$ we are denoting the determinant of 
$M_{i}$.

\section{Cosmic strings}
$         $ 

It is widely known that the space-time generated by a static cosmic string is
 described by the metric

\begin{equation}
\label{534}
ds^{2}=-dt^{2}+dz^{2}+d\rho^{2}+a^{2}\rho^{2}d\theta^{2},
\end{equation}
where $-\infty <t,z<\infty$, $0<\rho<\infty$, $0\leq\theta<2\pi$. The 
metric of this space-time, which possesses a 2-dimensional submanifold 
with a conical singularity, clearly has the form (\ref{432}). As is
 also evident from (\ref{432}), one of the 2-dimensional manifolds is 
readily identified with the plane $\Re^{2}$ while the other
 is the cone, the curvature scalar of which was calculated in section III.
 Thus, from the definition of $(G^{\lambda}_{\mu})^{*}$ and taking into 
account (\ref{330}) we end up with

\begin{equation}
\label{535}
G^{t}_{t}=G^{z}_{z}=-2\pi(1-a)\delta^{(2)}(\rho),
\end{equation}
while all the other components vanish.

If we assume the validity of the Einstein equations $G^{\lambda}_
{\mu}=8\pi GT^{\lambda}_{\mu}$ ($G$ is the gravitational constant and 
units are chosen in which $c=1$), then we determine the energy-momentum 
tensor $T^{\lambda}_{\mu}$ of the material source that generates the
 gravitational field described by the metric (\ref{534}):

\begin{equation}
\label{536}
T^{t}_{t}=T^{z}_{z}=\frac{(1-a)}{4G}\delta^{(2)}(\rho),
\end{equation}
with all the other components equal to zero. From this result we conclude 
that the matter source is concentrated on the axis $\rho=0$ with a linear 
mass density $\mu=\frac{(1-a)}{4G}$. Such configuration of matter has 
exactly the same structure as a vacuum string \cite{Vilenkin}, a topological 
defect predicted by gauge theories with spontaneous symmetry breaking. 
Actually, it was Sokolov and Starobinskii \cite{Sokolov} who first showed the connection between 
the space-time (\ref{534}) and the material source described by the 
energy-momentum tensor (\ref{536}). Curiously enough
 the same solution was rediscovery by Vilenkin \cite{Vilenkin} (also by Hiscock
 \cite{Hiscock} and Gott \cite{Gott} later) to a certain extent by following the
 inverse path, i.e., starting from (\ref{536}), solving the Einstein
 equations and, then, arriving at (\ref{432}).

\section{Final remarks}
$         $ 

By applying some concepts and definitions of distribution theory we have
 developed a formalism which may be applied to calculate, or more 
precisely, to define the Riemannian curvature of a certain class of 
space-times wherein conical singularities appear. It is known that the 
treatment of general space-times using distribution theory  
has revealed to be rather problematic. One of the main 
difficulties lies on the non-linearity of the Riemann tensor with respect 
to the affine connections. An attempt to formulate General Relativity 
theory in terms of distribution theory combining both mathematical rigour 
and high degree of generality has been undertaken by Geroch and Traschen 
\cite{Geroch}, who have shown that in arbitrary space-time a product of 
connections do not make sense as a distribution unless the connections 
satisfy some specific mathematical properties,e.g., they must be locally 
square-integrable. From to this last condition as well as  the peculiar way how the Riemann tensor is written in terms of the connections Geroch and Traschen  go further  to demonstrate that singular distributions 
cannot have support on a submanifold of less then three dimensions.   This does not seem to be a desirable situation as a 
number of space-times, such as  the conical space-times, are left out 
of consideration. In this article we have tried to overcome 
this difficulty. We have shown that if we renounce the idea of generality we can work out a prescription to treat a class of space-times with conical 
singularities using distribution theory concepts. This is due to the fact that 
for the  particular class of space-times that we have considered one 
 can express the Riemann tensor 
in a form which avoids terms containing products  of distributions.

The extension of our method to more general space-times than those taken 
into account in (\ref{432}) in order to include, for example, general static 
and axially symmetric space-times  is currently under investigation.

Finally, if the underlying theory of gravity is not General Relativity 
the problem of how to obtain the energy-momentum tensor may get a bit more 
complicated. This is the case, for example, of the so-called scalar-tensor 
theories of which Brans-Dicke theory is a particular case \cite{Brans}. Space-
times with conical singularity have been found in the context of Brans-Dicke 
theory of gravity \cite{Barros}. Here an extra difficulty arises due to the fact 
that the energy-momentum tensor is not determined by the geometry solely but 
depends also upon the scalar field.\\
$    $\\
$    $\\
{\bf Acknowledgements}\\
$     $\\
$     $\\
We are indebted to Prof. P. S. Letelier for helpful discussions. The authors 
thanks CNPq (Brazil) for financial support.\\
$     $\\
$     $\\
{\bf Appendix}\\
$     $

The  theorem mentioned in section III states the following:\\
Let us be given two real functions $g(s)$ and $h(s)$ which are continuous 
in the interval $L$: $a\leq s\leq b$. Assuming that $h(s)$ does not change
 its sign in $L$, then there exists a number $c$ lying on $L$ such that 

\begin{eqnarray}
\int_{a}^{b}g(s)h(s)ds=g(c)\int_{a}^{b}h(s)ds.\nonumber
\end{eqnarray}
Thus, chosing $g(s)$ and $h(s)$ as $\varphi(s)$ and $\frac{d\chi}{ds}$, respectively, and 
taking into account that $\frac{d\chi}{ds}=-\left[\frac{D\stackrel{\wedge}{e}_{x}}{Ds}
\right]=-\frac{1}{\epsilon}
\left[1-\frac{1}{\lambda^{2}(1+a^{2}\cos^{2}\left(\frac{s}
{\lambda\epsilon}\right))}\right]\geq 0$ , we have

\begin{eqnarray}
\int_{\partial S_{\epsilon}}\frac{d\chi}{ds}
\varphi(s)ds=
\varphi(c)\int_{\partial S_{\epsilon}}\frac{d\chi}{ds}ds,
\nonumber
\end{eqnarray}
where $c\in [0,2\pi\lambda\epsilon]$. It may happen that the number 
$c$ depends on $\partial S_{\epsilon}$, i.e., $c=c(\epsilon)$; 
nevertheless due to the continuity of the function $\varphi$ we must 
have 

\begin{eqnarray}
\lim_{\epsilon\rightarrow 0}\varphi(c)=\varphi(0).\nonumber
\end{eqnarray}


\begin{references}
 
\bibitem{Vilenkin} A. Vilenkin,  Phys. Rev. D {\bf 23}, 852 (1981).

\bibitem{Marder} L. Marder,  Proc. R. Soc. {\bf 252},  45 (1959).

\bibitem{Sokolov} D. D. Sokolov,  and A. A. Starobinskii, , Sov. Phys. 
Dokl. {\bf 22} (6), 312 (1977).

\bibitem{Kibble} T. W. B. Kibble,  J. Phys. A {\bf 9}, 1387 (1976).

\bibitem{Hiscock} W. A. Hiscock,  Phys. Rev. D {\bf 31}, 3288 (1985).

\bibitem{Gott} J. R. Gott III,  Astrophys. J. {\bf 288}, 422 (1985).

\bibitem{Linet} B. Linet,  Gen. Rel. Grav. {\bf 17}, 1109 (1985).

\bibitem{Israel} W. Israel,  Phys. Rev. D {\bf 15}, 935 (1977).

\bibitem{Fursaev} D. V. Fursaev,  and S. N. Solodukhin,  Phys. Rev. D {\bf 52},
 2133 (1995).

\bibitem{Clark} C. J. S. Clarke, J. A. Vickers,  and J. P. Wilson,  Class Quantum Grav. {\bf 13}, 2485 (1996).

\bibitem{Lich} A. Lichnerowicz,  C. R. Acad. Sci. {\bf 273}, 528 (1971).

\bibitem{Israel2} W. Israel,  Nuovo Cimento {\bf 44B}, 1 (1966).

\bibitem{Taub} A. Taub,  J. Math. Phys. {\bf 21}, 1423 (1980).

\bibitem{Letelier} P. S. Letelier  and A. Wang,  J. Math. Phys. {\bf 36}, 3023 
(1995). P. S. Letelier, J. Class. Quantum Grav. {\bf 4}, L75
(1987).

\bibitem{Geroch} R. Geroch,  and J. Traschen,  Phys. Rev. D {\bf 36}, 1017 
(1987).

\bibitem{Guelfand} I. M. Guelfand  and G. E. Chilov,  {\it Les distributions},
 (Dunod, Paris, 1962).

\bibitem{Carmo} M. P. do Carmo,  {\it Differential Geometry of Curves and
Surfaces} (Prentice-Hall, New Jersey, 1976).

\bibitem{Weinberg} S. Weinberg,  {\it Gravitation and Cosmology: Principles 
and Applications of the General Theory of Relativity} (John Wiley and 
Sons, New York, 1972). 

\bibitem{Eisenhart} P. L. Eisenhart,  {\it Riemannian Geometry} (Princeton 
University Press, Princeton, 1949).

\bibitem{Brans} C. Brans and R. H. Dicke, Phys. Rev. {\bf 124},
 925 (1961).

\bibitem{Barros} A. Barros and C. Romero,  J. Math. Phys. {\bf 36}, 5800 
(1995).
 
\end{references}
\end{document}